\documentclass{article}
\usepackage{arxiv}
\usepackage{amsmath,amssymb}
\usepackage{graphicx}
\graphicspath{{figures/}}
\usepackage{booktabs}
\usepackage{natbib}
\usepackage{microtype}
\usepackage{algorithm}
\usepackage{algpseudocode}
\usepackage{url}

\title{Neural Posterior Estimation for Stochastic Epidemic Models Using Final Outcome Data}
\author{Theodore Kypraios \\ School of Mathematical Sciences \\ University of Nottingham}
\renewcommand{\headeright}{}
\renewcommand{\undertitle}{}
\renewcommand{\shorttitle}{NPE for Epidemic Model Using Final Size Data}

\begin{document}

\maketitle
\begin{abstract}
    Neural posterior estimation (NPE) is a simulation-based approach to Bayesian inference that trains a neural network to approximate the posterior distribution from simulated parameter -- data pairs, bypassing likelihood evaluation.  We apply NPE --- to our knowledge for the first time --- to stochastic susceptible-infectious-removed (SIR) epidemic models observed through final outcome data, considering both homogeneously mixing and household-structured populations. Such data arise naturally in     retrospective outbreak investigations and household transmission studies, yet inference is computationally challenging: data-augmentation Markov chain Monte Carlo (MCMC) can be slow to mix in large populations and difficult to implement, while Approximate Bayesian Computation (ABC) suffers from low acceptance rates, particularly for large populations or unlikely outcomes. The discrete, low-dimensional nature of such observations makes this setting particularly well suited to NPE. We show that a logNormal posterior approximation, parameterised by a feed-forward neural network, accurately recovers reference posteriors across a range of population sizes and transmission regimes, and extends naturally to joint inference on global and local transmission rates in the household model. Once trained, the network produces approximate posterior distributions in seconds and generalises reliably to population sizes and structures not seen during training. Performance on both synthetic and real outbreak datasets is consistently strong, with results in close agreement with published analyses.
\end{abstract}

\keywords{Infectious disease epidemiology \and final outcome data \and Bayesian inference \and simulation-based inference}

\section{Introduction}\label{sec:intro}
Mathematical models of infectious disease transmission continue to play a central role in understanding epidemic dynamics, quantifying transmissibility, and informing public health decision-making. Ideally, during the course of an outbreak, one would collect detailed data at fine temporal resolution, such as individual-level symptom onset and recovery times. In many settings, however, such detailed temporal data are difficult to obtain or entirely unavailable. Moreover, the exact times at which individuals become infected are inherently unobserved. 

As a result, statistical inference must often rely on {\em final outcome data}, for example, the total number of infections observed in a closed population, the distribution of outbreak sizes across households, or serological surveys measuring cumulative infection incidence after an epidemic wave. Such data arise naturally in retrospective outbreak investigations, household transmission studies, and seroprevalence surveys conducted following epidemic peaks.  Developing reliable and efficient statistical methods for fitting stochastic epidemic models to these forms of aggregated data is therefore of substantial practical importance. The importance of such methods is underscored by continued 
research activity in this area: \citet{Brooks_etal_2026}, for example, recently developed Bayesian inference methods for household transmission data at varying levels of observational resolution, motivated in part by the challenges of analysing household COVID-19 transmission studies. Throughout this paper, \emph{final size} refers to the scalar total infection count in a homogeneously mixing population, and \emph{final outcome} refers more generally to the distribution of infection counts across households; we use \emph{final outcome data} as the collective term where the distinction need not be drawn.

The stochastic  {\em Susceptible–Infectious–Removed} (SIR) model in a homogeneously mixing population has served as a foundational building block of mathematical epidemiology \citep[see, for example,][and the references therein]{AnderssonBritton_2012}, and the final size of such an epidemic has been a central object of study both for model analysis and statistical inference \citep{Ball_1986}. A substantial body of work has extended this framework to incorporate population structure. Independent household models \citep{Addy_etal_1991, becker_dietz_1995} assume that within-household transmission proceeds as a standard SIR epidemic while between-household infections are modelled independently, yielding tractable likelihoods at the cost of approximating the true dependence structure. The so-called {\em two-level mixing model} introduced by \citet{Ball_etal1997} relaxes this, by explicitly modelling global transmission across the whole population and local transmission within groups, capturing the dependence structure more faithfully but rendering the likelihood intractable. Further extensions to random network and social cluster models provide additional flexibility in representing realistic contact patterns \citep[see e.g.][]{AnderssonBritton_2012}. 

Statistical inference for stochastic epidemic models observed through final outcome data has been approached in several ways, each with practical limitations. For the homogeneously mixing SIR model, the final size distribution can in principle be evaluated via a triangular system of recursive equations \citep{Ball_1986}, enabling direct likelihood computation. However, the bimodal nature of the distribution causes severe numerical instabilities even for moderate population sizes. \citet{DemirisONeill_2006} proposed the use of multiple-precision arithmetic to stabilise these calculations, though this requires specialised implementation and becomes increasingly costly as population size grows. For structured population models such as the two-level mixing model, exact likelihood evaluation is generally infeasible, and a common practical simplification is to assume independence between households, which is asymptotically justified as the number of groups grows but introduces approximation error of uncertain magnitude in finite populations \citep{Addy_etal_1991}.  Within a Bayesian framework, the most principled approaches avoid such approximations either by augmenting the latent transmission structure via MCMC \citep{DemirisONeill_2005, ONeill_2009} or by bypassing likelihood evaluation entirely through Approximate Bayesian Computation \citep{Kypraios_etal_2017, Neal_2012, ClancyOneill_2007}. Both strategies have been successfully applied to epidemic models of the kind considered here, but each carries a computational cost that grows with population size and becomes more acute as population structure increases in complexity.

Recent advances in simulation-based inference (SBI) \citep{Cranmer_etal_2020} offer a promising alternative. Rather than evaluating the likelihood directly or augmenting unobserved transmission events, these methods rely on repeated simulation from the model to learn an approximation to the posterior distribution of its parameters given observed data. In particular, neural posterior estimation (NPE) uses flexible density estimators trained on simulated data to approximate the mapping from observed data to the posterior distribution. Once trained, such models can provide rapid posterior inference for new datasets, offering the potential for substantial computational gains through amortisation.

In this paper, we investigate the use of NPE for Bayesian inference in stochastic epidemic models using final outcome data. We focus on the SIR model in both single-population and household-structured settings, representing two widely studied but computationally distinct scenarios. Through extensive simulation studies across a range of population sizes and transmission regimes, we assess the accuracy and computational efficiency of the proposed approach. We compare its performance with established Bayesian methods and examine the empirical structure of the resulting posterior distributions. Our results demonstrate that simulation-trained approximate posteriors can recover accurate approximation to the true posteriors while substantially reducing computational cost, suggesting a viable and scalable alternative to traditional data-augmentation approaches for epidemic modelling.

The remainder of the paper is organised as follows. Section~\ref{sec:models} introduces the two epidemic models and their associated inferential challenges. Section~\ref{sec:SBI} describes the NPE framework, illustrated through the homogeneously mixing model. Sections~\ref{sec:homo-SIR} and~\ref{sec:household-SIR} present the methodology and results for the homogeneous and household settings respectively, including applications to real outbreak data. Section~\ref{sec:discussion} concludes with a discussion of practical considerations and future directions.

\section{Models}\label{sec:models}
\subsection{Homogeneously mixing SIR model}\label{sec:homo-SIR-def}
\subsubsection{Model Definition}
We begin by describing the stochastic SIR model in a homogeneously mixing population. Consider a closed population of size $N$. A single individual is infected from outside the population and initiates an epidemic, with no subsequent external introductions. The remaining $N-1$ individuals are initially susceptible. Infectious individuals have independent and identically distributed infectious periods given by a random variable $I$, assumed to follow a specified parametric family. While infectious, individuals make infectious contacts according to a homogeneous Poisson process with rate $\beta$. Each contact is directed to an individual chosen uniformly at random from the population. If the contacted individual is susceptible, they become infected and immediately infectious; contacts with individuals who are not susceptible have no effect. At the end of their infectious period, individuals move to the removed class, corresponding to recovery with immunity or, in some cases, death, and play no further role in transmission. The epidemic terminates when no infectious individuals remain, and the total number of removed individuals at this time defines the final size of the epidemic.

\subsubsection{Simulation} \label{sec:homo-SIR-sim}
When inference is based solely on the final size of the epidemic, it is unnecessary to simulate the full temporal trajectory of infection and recovery events. We use the Sellke construction \citep{Sellke_1983}, which provides an efficient way to determine the final size directly: each susceptible individual is assigned a random infection threshold, and becomes infected once the cumulative infectious pressure generated by infectives  exceeds their threshold. Following \citet{Sellke_1983} and \citet{Neal_2012}, the final size $M_\beta$ is given by

\begin{equation} \label{eq:coupled:1}
M_\beta = \min \left\{ m : \sum_{i=1}^m L_i > \beta \sum_{j=1}^m I_j \right\},
\end{equation}

where $I_1,\ldots,I_N$ are independent draws from the infectious period distribution and $L_1,\ldots,L_{N-1}$ are independent with $L_i \sim \mathrm{Exp}\!\left((N-i)/N\right)$. Algorithm~\ref{alg:simSIR_final_size} summarises the procedure; we refer the reader to \citet{Neal_2012} and \citet{Kypraios_etal_2017} for full details.

\begin{algorithm}[H]
{\bf Input:} population size $N$, transmission rate $\beta$, 
             parameters of the infectious period distribution $f(I)$\\
{\bf Output:} final size $M_\beta$
\begin{algorithmic}[1]
\State Simulate $I_1, I_2, \ldots, I_N$ independently from $f(I)$.
\State Simulate $L_1, L_2, \ldots, L_{N-1}$ independently with 
       $L_i \sim \mathrm{Exp}\!\left(\frac{N-i}{N}\right)$ for 
       $i=1,\ldots,N-1$.
\State Compute $T_{(i)} = \sum_{j=1}^{i-1} L_j$ for $i=1,\ldots,N$.
\State Determine $M_\beta$ using Equation~(\ref{eq:coupled:1}).
\end{algorithmic}
\caption{Simulation of the final size of a homogeneously mixing SIR 
         epidemic using the Sellke construction.}
\label{alg:simSIR_final_size}
\end{algorithm}

\subsubsection{Inference using final size data} \label{sec:homo-SIR-inf}

For the homogeneously mixing SIR model observed through its final size, a number of inferential approaches have been developed. In principle, for a given infection rate, the probability mass function of the final size distribution can be evaluated using the triangular system of equations \citep{Ball_1986}. This allows exact computation of the observed data likelihood and hence enables both maximum likelihood and Bayesian inference. However, it is well known \citep[e.g.][]{AnderssonBritton_2012} that the recursive scheme can suffer from numerical instability due to the bimodal nature of the final size distribution. In particular, probabilities of interest may depend on intermediate quantities that are extremely small and fall below machine precision. These difficulties arise even for moderately sized populations and become more pronounced when exploring extreme regions of the parameter space during likelihood-based inference, limiting the practical applicability of direct recursive evaluation. To mitigate these issues, \cite{DemirisONeill_2006} proposed the use of multiple-precision arithmetic to stabilise likelihood calculations and facilitate Bayesian inference. Although this extends the range of tractable population sizes, it requires specialised numerical implementation and is not readily accessible within standard statistical software environments.

A different non-approximate strategy is based on data-augmentation using random directed graphs \citep{DemirisONeill_2005}. The central idea is to impute a detailed latent transmission structure describing, for each infected individual, the set of susceptible individuals that they would infect in the absence of competing infections. This structure can be represented as a random digraph whose edges correspond to potential infectious contacts. Conditional on the augmented graph, the likelihood becomes tractable and posterior inference can proceed via MCMC without reliance on independence approximations or recursive likelihood evaluation. While exact in the Monte Carlo sense and numerically stable, this approach introduces a high-dimensional latent object and requires careful MCMC design. As population size increases, the dimension of the augmented graph grows rapidly, leading to increased computational cost and potential mixing difficulties.

Approximate Bayesian Computation (ABC) \citep{Sisson_etal_2018} provides a likelihood-free alternative. In the homogeneously mixing SIR model, the final size is a one-dimensional sufficient statistic, so ABC can be implemented efficiently and, in some settings, even exactly; see \cite{Kypraios_etal_2017}. More broadly, ABC has been widely applied to epidemic models with intractable likelihoods \citep{Neal_2012,MinterRetkute_2019,Mckinley_etal_2009}. However, ABC can be computationally intensive. Rejection-based schemes accept parameter draws only when simulated data closely match the observed final size, and acceptance rates may be extremely low for large populations or unlikely outcomes. Sequential and adaptive variants improve efficiency but still require substantial forward simulation and careful tuning, and may introduce additional approximation error when exact matching is not achievable.

\subsection{Household SIR epidemic model}
\subsubsection{Model Definition} \label{sec:two-level-def}
An important and widely studied extension of the homogeneously mixing model (defined in Section \ref{sec:homo-SIR-def}) is the household epidemic model; see \cite{Ball_etal1997}. We consider a closed population of $N$ individuals partitioned into $H$ households. Let $K$ denote the maximum household size and, for $k=1,2, \ldots, K$, let $H_k$ represent the number of households of size $k$. Define $H = \sum_{k=1}^{K}H_k$ and $N = \sum_{k=1}^{K} k\,H_k$.  As in the homogeneously mixing model, individuals who become infected have independent and identically distributed infectious periods, denoted by the random variable $I$. During their infectious period, individuals may make infectious contacts with any member of the population, but transmission is structured to allow for elevated contact rates within households.

Specifically, infectious individuals generate global infectious contacts according to a homogeneous Poisson process with rate $\lambda_G$, with each contact directed uniformly at random to a member of the entire population. In addition, while infectious, an individual $i$ makes infectious contact with each other member of their household according to a homogeneous Poisson process with rate $\lambda_L$, corresponding to density-dependent transmission within households. When $\lambda_L = 0$, the model reduces to the homogeneously mixing epidemic described previously.

Both models above do not include a latent period. However, including a latent period makes no difference to the distribution of the final outcome of the epidemic. Therefore, the methods described in this paper can be applied to epidemic models that include a latent period.

\subsubsection{Simulation} \label{sec:two-level-sim}
For simulation of the household epidemic model we employ a Sellke-type construction \citep{Sellke_1983}, adapted to the two-level mixing structure following \citet{Ball_etal1997}. Since we require only the final outcome, it is not necessary to simulate the full temporal evolution of the epidemic. The key idea is to separate the epidemic into successive global infections and the local within-household outbreaks that each one initiates.

An individual is selected uniformly at random as the initial infective. This individual triggers a within-household outbreak of size $Y_1$ (including the initial case) and total severity $S_1$, where severity is defined as the sum of infectious periods of all individuals infected in that outbreak. For $j = 1, 2, \ldots$, let $(Y_j, S_j)$ denote the size and severity of the local epidemic generated by the $j$th global infection. Conditional on the household size and the number of susceptibles present at the start of the $j$th local outbreak, the pairs $\{(Y_j, S_j)\}$ are independent, and each can be simulated using Algorithm~\ref{alg:simSIR_final_size} with minor modifications.

Global infections are governed by the threshold mechanism
\begin{equation}
L_i \sim \mathrm{Exp}\!\left(\frac{N - \sum_{j=1}^{i} Y_j}{N}\right),
\label{eq:house:1}
\end{equation}
where $L_i$ is the additional global infectious pressure needed, after the $i$th global infection, to trigger the $(i+1)$st, accounting for the depletion of susceptibles by the first $i$ local outbreaks. The total number of global infections is then
\begin{equation}
M_{\lambda_G} = \min\left\{m : \sum_{i=1}^{m} L_i > \lambda_G 
              \sum_{j=1}^{m} S_j\right\}.
\label{eq:house:2}
\end{equation}
The individuals ultimately infected are those involved in the first $M_{\lambda_G}$ global infections and their associated within-household outbreaks. We refer the reader to \citet{Ball_etal1997} and \citet{Neal_2012} for full details.

\subsubsection{Inference using final size data} \label{sec:two-level-inf}

For the household SIR model, the inferential objective is to estimate the local $(\lambda_L)$ and global $(\lambda_G)$ transmission rates given the population structure and final outcome data indicating which individuals were infected. Likelihood-based inference is challenging because global transmission induces dependence across households: infections in one household are not independent of those in others, so the likelihood cannot be factorised into a product over households. Consequently, exact evaluation of the observed-data likelihood is generally infeasible except for very small populations.

A common simplification is to assume independence between households, an approximation that is reasonable in large populations and holds asymptotically as the number of households grows. Under this assumption, the two-level mixing model reduces to an independent-households model in which local transmission proceeds as usual, but each individual independently avoids infection from outside the household with some fixed probability. This yields a tractable likelihood and enables inference using standard methods \citep[see, e.g.,][]{Addy_etal_1991}. However, this approach neglects the dependence structure induced by global transmission and is therefore only approximate.

More flexible, non-approximate approaches rely on Bayesian data-augmentation. The random digraph MCMC strategy of \cite{DemirisONeill_2005} extends naturally to the household setting, and related schemes \citep[e.g.][]{ONeill_2009} introduce additional latent contact information at the individual level. Conditional on the augmented transmission structure, the likelihood becomes tractable and inference can proceed without independence assumptions. Nevertheless, these methods involve exploring a substantially higher-dimensional latent state space than in the homogeneously mixing case. As the number and size of households increase, computational cost grows rapidly and MCMC algorithms may suffer from slow mixing and strong posterior dependence. Likelihood-free approaches such as ABC have also been applied \citep{Neal_2012}, but inference is considerably more demanding than in the homogeneous setting. The absence of low-dimensional sufficient statistics necessitates the construction of informative summary statistics, introducing additional approximation and potential information loss. Consequently, scalable and computationally efficient Bayesian inference for household epidemic models based solely on final outcome data remains challenging.

Taken together, existing approaches either rely on structural approximations, introduce high-dimensional latent variables with substantial computational overhead, or incur additional approximation through summary statistics. These challenges motivate the development of alternative simulation-based Bayesian methods that can exploit forward simulation of the model while avoiding explicit likelihood evaluation or high-dimensional data-augmentation.

\section{Simulation-Based Bayesian Inference} \label{sec:SBI}

\subsection{Overview of simulation-based inference}

Simulation-based inference (SBI) has emerged as a powerful and general class of methods for performing Bayesian inference in models for which the likelihood is unavailable or computationally intractable, but forward simulation from the model is feasible \citep{Cranmer_etal_2020}. Rather than evaluating the likelihood explicitly, SBI methods approximate the posterior distribution by repeatedly simulating synthetic datasets under the model and using these simulations to construct a surrogate for the posterior.  Formally, let $\theta$ denote model parameters and $y$ observed data. Even when the likelihood $\pi(y|\theta)$ cannot be evaluated pointwise, it is often possible to generate simulated data $y \sim \pi(y|\theta)$. SBI methods exploit this capability to approximate the posterior, $\pi(\theta|y) \propto \pi(y | \theta)\, \pi(\theta)$, using simulated pairs $\{(\theta_i, y_i)\}_{i=1}^M$ drawn from the prior and the forward/generative model.

Modern SBI methods can be broadly categorised according to which component of the Bayesian decomposition they approximate. Neural likelihood estimation  methods \citep{papamakarios2019sequential} learn a surrogate likelihood function; neural ratio estimation methods \citep{hermans2020likelihood} learn the likelihood-to-evidence ratio; and neural posterior estimation (NPE) methods learn the posterior distribution directly. In this work we focus on NPE, which aims to approximate $\pi(\theta | y)$ using flexible conditional density estimators trained on simulated parameter-data pairs. Conceptually, SBI can be viewed as a generalisation of ABC. Modern SBI methods retain the simulation-based philosophy of ABC but replace distance-based acceptance and regression adjustments with flexible neural density estimators that learn the full probabilistic relationship between parameters and data. In particular, NPE avoids the need to specify tolerance levels and can, in principle, operate directly on raw or minimally processed data without ad hoc summary construction.

The key idea underlying NPE and related approaches is to first (i) generate a training dataset consisting of parameter-data pairs simulated from the model, and then (ii) train a neural network, often referred to as an {\em inference network}, to learn a conditional density $q_\phi(\theta| y)$ that approximates the true posterior \citep{PapamakariosMurray_2016, Greenberg_etal_2019}. After training, the network is evaluated at the observed dataset to produce an approximation to the posterior distribution. Because inference relies solely on model simulations, these methods do not require explicit likelihood evaluations and are  applicable whenever forward simulation is possible. Moreover, many SBI approaches enable \emph{amortised inference}: once trained, the network can be reused to perform inference for new datasets without requiring additional simulations or retraining.

\begin{algorithm}[H]
{\bf Input:} number of pairs $M$, parameters governing $\pi(\theta)$\\
{\bf Output:} Samples $\{(\theta_i, y_i)\}_{i=1}^M$ drawn from $\pi(\theta, y)$
\begin{algorithmic}[1]
\For{$i=1, \ldots, M$}
\State draw $\theta_i \sim \pi(\theta)$;
\State draw $y_i \sim \pi(y|\theta_i)$
\EndFor
\end{algorithmic}
\caption{Generation of a training dataset}
\label{alg:SBI-generic}
\end{algorithm}

Much of the recent development of SBI has been driven by applications in fields such as cosmology, particle physics, and systems biology, where models may involve dozens or hundreds of parameters and extremely high-dimensional observations \citep[see, for example,][]{dax2021real, dax2023neural, boelts2022flexible}. In contrast, epidemic models based on final outcome data typically involve only a small number of transmission parameters and comparatively low-dimensional summaries. Furthermore, posterior distributions in such models are often unimodal and reasonably well behaved. Thus, epidemic models do not necessarily present the extreme inferential challenges that originally motivated SBI. Instead, the primary difficulty lies in the intractability of the likelihood induced by dependence structures or latent transmission processes. This setting therefore provides a useful and controlled environment for assessing whether modern neural simulation-based methods can offer computational or practical advantages even when the underlying inferential geometry is relatively simple.

\subsection{Illustration for the homogeneously mixing SIR model} \label{sec:homo-SIR-illust}
\begin{figure}[t]
    \centering
    \includegraphics[width=0.8\linewidth]{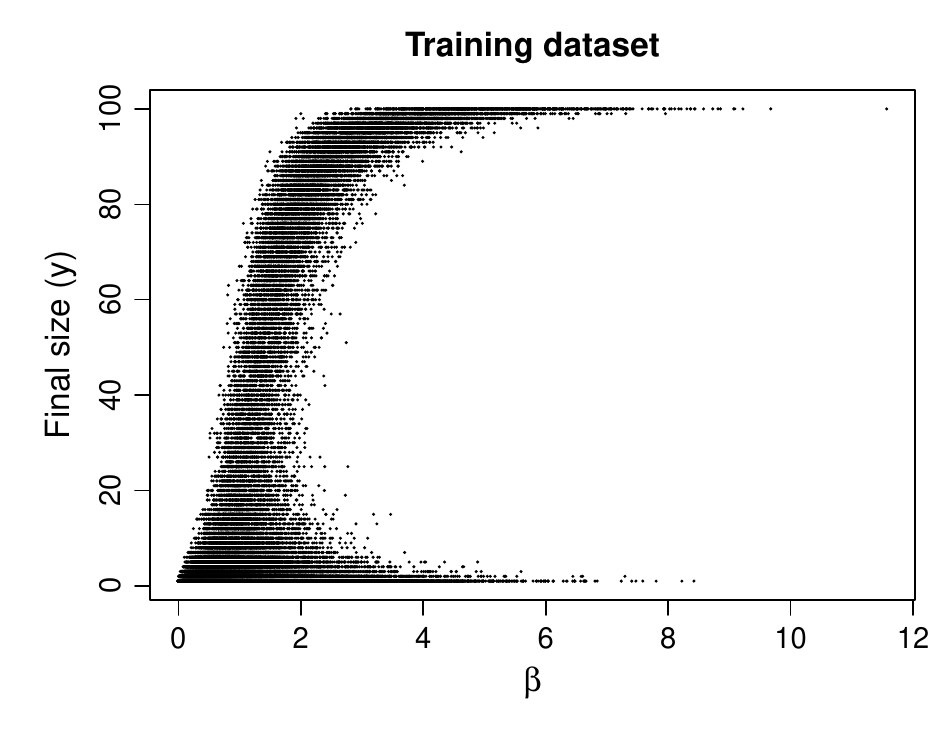}
    \caption{Illustration of simulated training data under the  homogeneously mixing SIR model with $N = 100$. Each point corresponds to a parameter--data pair $(\beta, y)$ generated by sampling $\beta$ from the prior and simulating the corresponding final size. The bimodal structure visible in the plot, with points clustering near $y = 0$ (minor outbreaks) and spreading across large $y$ (major outbreaks), reflects the well-known threshold behaviour of the SIR 
    model discussed in Section~\ref{sec:homo-SIR-inf}.}
    \label{fig:training-data-illustration}
\end{figure}

To illustrate the basic idea, we consider the homogeneously mixing SIR model (\ref{sec:homo-SIR-def}) and generate a training dataset using Algorithm \ref{alg:SBI-generic} by repeatedly sampling parameter values (e.g. infection rate) from the prior and simulating the corresponding final size. Figure~\ref{fig:training-data-illustration} shows the resulting collection of $10^5$ parameter–data pairs in a population of size $N=100$ individuals out of whom there was one initially infective and assuming an Exp(1) prior for the infection rate and an Exp(1) infectious period.  The stochastic nature of the epidemic process induces variability in final size even for fixed parameter values, resulting in a diffuse cloud rather than a deterministic curve. The training dataset therefore represents draws from the joint distribution $\pi(\theta, y)$ under the prior and the model.

Suppose now that the observed final size is $y_{\mathrm{obs}}$.%, indicated by a horizontal line in Figure~\ref{fig:training-data-illustration}. 
The posterior distribution $\pi(\theta|y_{\mathrm{obs}})$ corresponds to the conditional distribution of $\theta$ given $y = y_{\mathrm{obs}}$. Conceptually, this can be thought of as looking at a ``cut'' through the joint density $\pi(\theta, y)$ at $y_{\mathrm{obs}}$ and inspecting the distribution of corresponding $\theta$ values.  In practice, however, obtaining the posterior from such a ``cut'' poses a fundamental challenge: for continuous data, no simulation will exactly match the observed value $y_{\mathrm{obs}}$. Furthermore, even in the case of discrete data, there may not be enough samples $\{\theta_i, y_i\}$ for which $y_i = y_{\mathrm{obs}}$. ABC addresses this challenge by retaining those parameter values whose simulated final sizes lie within a tolerance region around $y_{\mathrm{obs}}$. In other words, ABC performs a local filtering of the simulated cloud based on proximity in the data space.

In contrast, simulation-based inference methods such as NPE use the entire simulated dataset to learn a smooth approximation to the conditional distribution $\pi(\theta | y)$. Rather than discarding simulations outside a tolerance region, all simulated pairs contribute to the estimation of a global conditional density model. Once trained, this model can be evaluated directly at $y_{\mathrm{obs}}$ to produce an approximation to the posterior distribution. We explain how this is done in the following section.

\subsection{Neural Posterior Estimation}

Neural posterior estimation seeks to approximate the posterior distribution $\pi(\theta | y)$ by a parametric conditional density $q_\phi(\theta | y)$, where $\phi$ denotes a set of unknown parameters to be estimated by minimising a discrepancy between $\pi(\theta | y)$ and $q_\phi(\theta | y)$.  A natural choice of discrepancy is the Kullback–Leibler (KL) divergence, $\mathrm{KL}\big(\pi(\theta | y) \,\|\, q_\phi(\theta | y)\big)$. We seek $q_\phi(\theta | y)$ to be close to $\pi(\theta | y)$, on average, over datasets $y$ drawn from the prior predictive distribution. This leads to the loss function
$\tilde{\mathcal{L}}(\phi) = \mathbb{E}_{\pi(y)}\!\left[\mathrm{KL}\big(\pi(\theta | y)\,\|\, q_\phi(\theta | y)\big)\right]$. It is straightforward to show that minimising  $\tilde{\mathcal{L}}(\phi)$ is equivalent to minimising
\[
\mathcal{L}(\phi) 
= 
\mathbb{E}_{(\theta, y) \sim \pi(\theta, y)}
\left[
- \log q_\phi(\theta | y)
\right],
\]
where $(\theta, y)$ are simulated by sampling $\theta \sim \pi(\theta)$ from the prior and $y \sim \pi(y | \theta)$ from the model (see Algorithm~\ref{alg:SBI-generic}). The reason is that expanding the KL divergence gives $\tilde{\mathcal{L}}(\phi) = \mathcal{L}(\phi) - \mathbb{E}_{\pi(y)}\bigl[\mathrm{H}\bigl(\pi(\theta|y)\bigr)\bigr]$, 
where $\mathrm{H}\bigl(\pi(\theta|y)\bigr)$ denotes the entropy of the 
true posterior. Since this entropy term does not depend on $\phi$, it plays no 
role in the optimisation, and minimising $\tilde{\mathcal{L}}(\phi)$ 
reduces to minimising $\mathcal{L}(\phi)$. In practice, this expectation is approximated using Monte Carlo simulation, and estimation therefore reduces to maximising the conditional log-density of $\theta$ under the model $q_\phi(\cdot | \cdot)$ given simulated data $y$. Training thus amounts to optimising the neural network weights $\phi$ with respect to this objective.

For illustration, suppose that for a given model (e.g. the homogeneously mixing SIR model) the posterior distribution of a scalar parameter $\theta$ (e.g. the infection rate) can be reasonably approximated by a logNormal distribution:
\[
q_\phi(\theta | y) 
\equiv 
\mathrm{logNormal}\big(\mu_\phi(y), \sigma_\phi^2(y)\big),
\]
where $\mu_\phi(y)$ and $\sigma_\phi^2(y)$ are the mean and variance of $\log\theta$ under the approximating distribution, each expressed as a  function of the data $y$ via the neural network parameters $\phi$. 

The inferential task then becomes one of learning suitable functional forms for $\mu_\phi(\cdot)$ and $\sigma_\phi(\cdot)$ and parameters $\phi$ using simulated parameter–data pairs. In principle, these functions could be specified using simple parametric regressions. However, to allow greater flexibility, modern approaches employ neural networks as universal function approximators. The network takes $y$ as input and outputs the parameters of the conditional density $q_\phi(\theta | y)$ (here, $\mu_\phi(y)$ and $\sigma_\phi(y)$). In this case, $\phi$ corresponds to the collection of weights and biases of the neural network with a chosen architecture, which effectively parameterises the conditional density model. More flexible density families can also be used, such as mixture density networks \citep{bishop1994mixture}, which represent $q_\phi$ as a finite mixture of simple component distributions  and can accommodate skewness or multimodality where required.

In the epidemic models considered in this paper, the parameter dimension is low and posterior distributions are often unimodal and relatively well behaved. Consequently, highly complex density estimators may not be necessary. Nevertheless, the NPE framework provides a principled and computationally efficient way to approximate the posterior without requiring likelihood evaluation, while retaining the flexibility to model more complex posterior structures if they arise.

A distinctive feature of neural posterior estimation is \emph{amortisation}. 
The neural network is trained once using simulated parameter–data pairs drawn from the prior predictive distribution, thereby learning a mapping from data $y$ to an approximate posterior density $q_\phi(\theta|y)$. 
After this training phase, inference for a new observed dataset requires only a forward pass through the network, without further simulation or optimisation. This differs from traditional approaches such as MCMC or ABC, where inference must be repeated separately for each dataset and often requires careful, model-specific algorithm design or tuning to ensure adequate performance. 
In contrast, NPE relies on a general-purpose training procedure that does not need to be redesigned for each new dataset within the same model class. In epidemic settings, where multiple outbreaks may be analysed or inference may need to be updated as data accumulate, this amortised structure can offer substantial computational and practical advantages.

\section{Homogeneously mixing SIR model} \label{sec:homo-SIR}
\subsection{Experimental setup}

We first consider inference for the infection rate parameter $\beta$ of the homogeneously mixing SIR model introduced in Section~\ref{sec:homo-SIR-def} based on observing the epidemic's final size. We assign $\beta$ an $\mathrm{Exp}(1)$ prior, reflecting positivity and concentrating prior mass in the epidemiologically relevant range of transmission intensities; the rationale for this choice, and its implications for training, are discussed in Section~\ref{sec:discussion}.

We approximate the posterior distribution with $q_\phi(\beta|y)$. We use a logNormal distribution, ensuring support on the positive real line. The logNormal parameters are modelled as functions of the data $y$ via a feed-forward neural network of a fixed architecture with three hidden layers of 64 units each and ReLU activation functions. The network is trained by minimising the negative log-likelihood of the approximate posterior over simulated parameter--data pairs (see Algorithm \ref{alg:SBI-generic}) using the Adam optimiser with a learning rate of $10^{-3}$, a batch size of 1024, and 100 epochs. The architecture is intentionally modest: the low dimensionality of the parameter space and the well-behaved nature of the posterior distributions in such a setting mean that a lightweight network is sufficient, and no regularisation or specialised training procedures are required.

To assess performance, we compare the NPE approximate posterior against  true  posterior obtained via rejection ABC with exact matching on the final size. In the homogeneously mixing model, where the data are one-dimensional and simulation is relatively inexpensive, at least for moderate population size $N$, such an approach provides a reliable approximation to the true posterior distribution. This comparison allows us to evaluate the accuracy of the learned posterior density in a controlled setting.

\subsection{Synthetic data experiments}

\subsubsection{Fixed population size $N$} \label{sec:npe_homSIR_fixed_N}
\begin{figure}[htbp]
    \centering
    \begin{minipage}{0.49\linewidth}
        \centering
        \includegraphics[width=\linewidth]{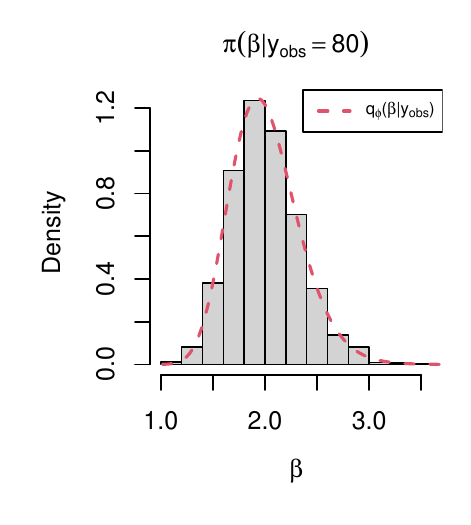}
    \end{minipage}
    \hfill
    \begin{minipage}{0.49\linewidth}
        \centering
        \includegraphics[width=\linewidth]{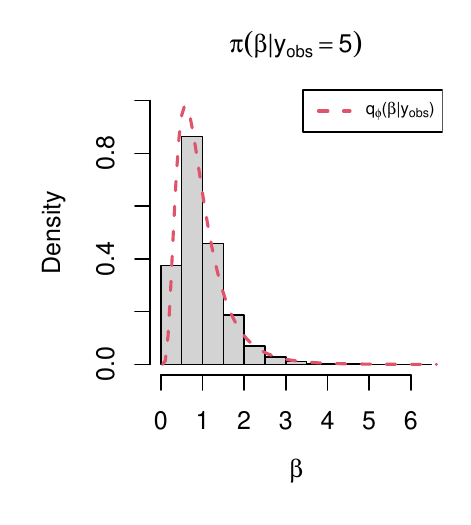}
    \end{minipage}
    \centering
    \begin{minipage}{0.49\linewidth}
        \centering
        \includegraphics[width=\linewidth]{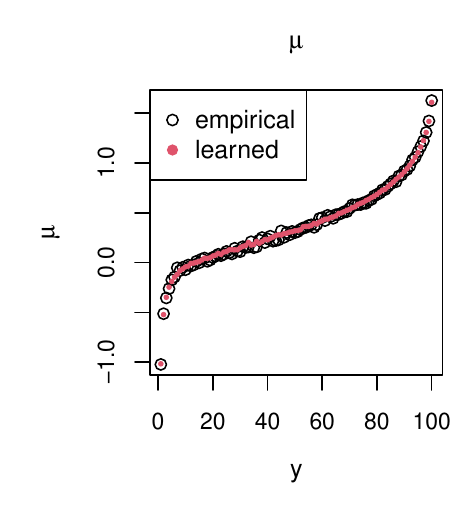}
    \end{minipage}
    \hfill
    \begin{minipage}{0.49\linewidth}
        \centering
        \includegraphics[width=\linewidth]{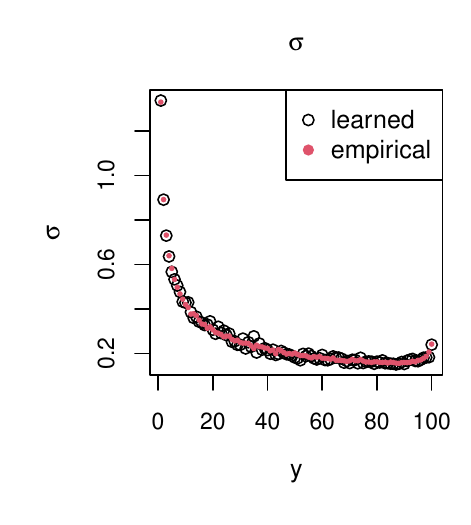}
    \end{minipage}
    \caption{Accuracy of the logNormal neural posterior estimator for the homogeneously mixing SIR model. The top row shows the reference and NPE posterior distributions for two contrasting observed final sizes ($y = 80$ and $y = 5$). The bottom row shows the learned functional relationship between $y$ and the logNormal approximation parameters: mean $\mu_\phi(y)$ (left) and standard deviation $\sigma_\phi(y)$ (right). Red solid circles show network outputs; white circles show maximum likelihood estimates obtained by fitting a logNormal distribution to simulated $\beta$ values conditional on each fixed $y$. The top row confirms accuracy at specific observations, while the bottom row confirms that the correct conditional posterior structure has been learned across the full observation range.}\label{fig:homo-SIR-NPE-N-100-functional-learning}    
\end{figure}

Recall the setting introduced in Section~\ref{sec:homo-SIR-illust}, where a closed population of size $N$ contains a single initial infective. The top row of Figure~\ref{fig:homo-SIR-NPE-N-100-functional-learning} compares the reference posterior distribution of the infection rate $\beta$ with the NPE approximation for two contrasting observed final sizes, $y=5$ and $y=80$. In both cases, the approximate posterior closely matches the reference distribution, indicating that the logNormal estimator provides an accurate description of the conditional posterior.

To assess whether the network has learned the structure visible in Figure~\ref{fig:training-data-illustration} globally, i.e. not merely for the two selected observations, the plots in the bottom row of Figure~\ref{fig:homo-SIR-NPE-N-100-functional-learning} examine the functional relationship between $y$ and the parameters of the logNormal approximation $\mu_\phi(y)$ and $\sigma_\phi(y)$, which are outputs of the neural network. Since $y$ takes values in $\{1,\ldots,N\}$, the trained network implicitly defines a posterior approximation for every possible final size in this range. We compare the learned functions $\mu_\phi(y)$ and $\sigma_\phi(y)$ with empirical estimates obtained directly from the simulated training data: for each fixed value of $y$, we extract all simulated infection rates $\beta$ that produced  that final size and fit a logNormal distribution by maximum likelihood, yielding empirical estimates of the conditional mean and standard deviation of $\log \theta$ as functions of $y$.

The graphs in the bottom row of Figure~\ref{fig:homo-SIR-NPE-N-100-functional-learning} show close agreement between the neural network predictions and these empirical estimates across the full range of feasible final sizes, indicating that the network has learned a smooth functional approximation to the entire conditional posterior family, rather than merely reproducing accurate approximations for selected datasets. This illustrates a key feature of the amortised approach: a single training run produces a coherent mapping from data to posterior for every admissible observation simultaneously, without any additional computation at the inference stage.

Table~\ref{tab:hom_sir_summaries} extends this assessment to three observed final sizes ($y = 25, 50, 75$) and three infectious period distributions---constant, Exponential, and Gamma---each with mean 1 and variances 0, 1, and 10 respectively, following the setup of \citet{DemirisONeill_2005}. Since the mean infectious period equals 1 in all cases, the basic reproduction number satisfies $R_0 = \beta$ numerically, and results are reported in terms of $R_0$ to facilitate epidemiological interpretation. For each combination we report posterior means and standard deviations from both the true posterior and NPE.

\begin{table}[ht]
\centering
\caption{Posterior means and standard deviations for $R_0$ under three infectious period distributions, each with mean 1 and variance 0 (constant), 1 (exponential), and 10 (gamma). Results are shown for the true posterior (obtained via exact matching ABC) and NPE across three observed final sizes, with population size $N = 100$ and a single initial infective.}
\label{tab:hom_sir_summaries}
\begin{tabular}{ll rr rr rr}
\toprule
& & \multicolumn{2}{c}{\textbf{Constant}} 
  & \multicolumn{2}{c}{\textbf{Exponential}} 
  & \multicolumn{2}{c}{\textbf{Gamma}} \\
\cmidrule(lr){3-4}\cmidrule(lr){5-6}\cmidrule(lr){7-8}
$y$ & & $\pi(R_0|y)$ & $q_{\phi}(R_0|y)$ 
    & $\pi(R_0|y)$ & $q_{\phi}(R_0|y)$  
    & $\pi(R_0|y)$ & $q_{\phi}(R_0|y)$  \\
\midrule
25  & Mean       & 1.141 & 1.123 & 1.151 & 1.170 & 1.257 & 1.280 \\
    & Std.\ dev. & 0.228 & 0.232 & 0.311 & 0.327 & 0.731 & 0.711 \\
\addlinespace
50  & Mean       & 1.401 & 1.394 & 1.419 & 1.398 & 1.461 & 1.472 \\
    & Std.\ dev. & 0.198 & 0.202 & 0.283 & 0.277 & 0.634 & 0.625 \\
\addlinespace
75  & Mean       & 1.859 & 1.845 & 1.866 & 1.866 & 1.811 & 1.840 \\
    & Std.\ dev. & 0.228 & 0.229 & 0.311 & 0.313 & 0.627 & 0.638 \\
\bottomrule
\end{tabular}
\end{table}

The agreement between NPE and the true posterior is excellent throughout. Posterior means match to within approximately 1--2\% across all combinations of final size and infectious period distribution, and standard deviations are similarly close. The results also exhibit the expected qualitative structure: posterior means increase with final size, reflecting the greater transmission intensity implied by larger outbreaks; and posterior uncertainty decreases as the infectious period becomes less variable, with the constant infectious period yielding the tightest posteriors and the gamma the widest. This pattern is consistent with the findings of \citet{DemirisONeill_2006}.

\subsubsection{Variable population size $N$} \label{sec:npe_homSIR_variable_N}
The network trained in Section~\ref{sec:npe_homSIR_fixed_N} was designed for a single, fixed population size $N$ and accepts only the final size $y$ as input. In practice, outbreaks occur in populations of varying size, and it would be inefficient to train a separate network for each $N$. A natural and practically useful extension is to include $N$ as an additional input to the network alongside $y$, so that a single trained model can perform inference across a range of population sizes without retraining.

We therefore augment the network input from $y$ to the pair $(y, N)$, while keeping the density estimator, loss function, and training procedure identical to those described in Section \ref{sec:npe_homSIR_fixed_N}. Training data are generated across $K = 5$ population sizes, $N \in \{100, 200, 500, 1000, 2000\}$, with $M = 10^6$ parameter--data pairs $(\beta, y)$ simulated for each value of $N$, giving a pooled training dataset of $5 \times 10^6$ triples $(\beta, y, N)$. To present both inputs on a comparable scale, the final size is expressed as the observed proportion infected, $y/N$, and the population size is normalised to $[0,1]$ by dividing by the largest value in the training set, $N_{\max} = 2000$. The network therefore learns the mapping $(y/N,\, N/N_{\max}) \mapsto \bigl(\mu_\phi(y, N),\, \sigma_\phi(y, N)\bigr)$, where the logNormal parameters now depend jointly on the observed proportion infected and the population size. The network architecture follows that of Section~\ref{sec:npe_homSIR_fixed_N}, with the exception that the hidden layer width is increased to 128 units and the number of training epochs to 200, reflecting the more demanding learning problem posed by the two-dimensional input $(y/N, N/N_{\max})$.

Figures \ref{fig:varN_mu} and \ref{fig:varN_sigma} in the Appendix show the learned logNormal parameters $\mu_\phi(y, N)$ and $\sigma_\phi(y, N)$ as functions of the final size $y$ for each of the five population sizes, alongside empirical estimates obtained by exact matching from the training data. The close agreement across all values of $N$ confirms that the single network has simultaneously learned the conditional posterior structure for each population size. As $N$ increases, the posterior becomes more concentrated (i.e. $\sigma_\phi(y, N)$ decreases with $N$ for a given proportion infected) which is consistent with the greater information content of a larger outbreak.

\begin{figure}[h]
    \centering
    \includegraphics[width=0.8\textwidth]{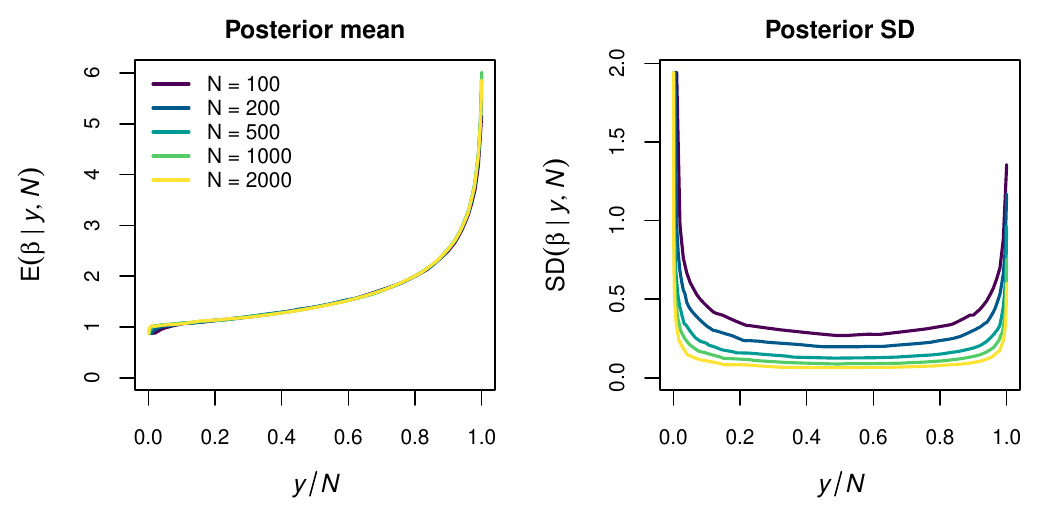}
    \caption{Posterior mean $\mathrm{E}(\beta|y, N)$ (left) and posterior standard deviation $\mathrm{SD}(\beta|y, N)$ (right) as functions of the proportion infected $y/N$, for each of the five training population sizes.
    Both quantities are derived from the logNormal parameters output by the trained network. The posterior mean is largely insensitive to $N$ for a given proportion infected, while the posterior standard deviation decreases
    markedly with $N$, reflecting the greater informativeness of larger outbreaks.}
    \label{fig:varN_mean_sd}
\end{figure}

Figure~\ref{fig:varN_mean_sd} summarises what the trained network has learned about the dependence of the posterior on both $y$ and $N$. The left panel shows the posterior mean $\mathrm{E}(\beta|y, N)$ as a function of the proportion infected $y/N$ for each of the five population sizes. The curves are strikingly similar across all values of $N$, indicating that the location of the posterior is governed primarily by the proportion of the population infected, rather than by the absolute population size. This is consistent with classical asymptotic results for the SIR model, in which the final size proportion converges to a deterministic limit as $N \to \infty$ \citep{AnderssonBritton_2012}. The right panel shows the posterior standard deviation $\sqrt{\mathrm{var}(\beta|y, N)}$, which, in contrast, decreases substantially with $N$ for any fixed $y/N$. This reflects a well-known feature of epidemic inference: larger populations provide more precise information about the transmission rate, since the stochastic variability in the final size becomes relatively smaller as $N$ increases. Taken together, the two panels illustrate that including $N$ as a network input allows the model to correctly capture this dependence: the posterior location is shared across population sizes, but the posterior uncertainty is appropriately modulated by $N$.

A key practical advantage of this formulation is that inference for a new outbreak requires only a single forward pass through the network, regardless of whether $N$ was seen during training or lies within the range covered. No additional simulations or retraining are required. This is especially valuable in settings such as household transmission studies or seroprevalence surveys, where population sizes vary across datasets and rapid repeated inference may be needed.

To demonstrate the practical value of the amortised variable-$N$ approach, we evaluate the trained network at two population sizes that were not included in the training data. The first case, $y_{\mathrm{obs}} = 240$ and $N = 300$, represents a major outbreak in which $80\%$ of the population was ultimately infected. The second, $y_{\mathrm{obs}} = 350$ and $N = 1500$, represents a considerably milder episode in a much larger population, with a proportion infected of approximately $0.23$. These two cases are deliberately chosen to occupy contrasting regions of the $(y/N, N)$ input space: one at high proportion infected and moderate $N$, the
other at low proportion infected and large $N$. In both cases, the network is required to interpolate between training population sizes rather than simply reproduce a learned relationship at a seen value of $N$.

\begin{figure}[htbp]
    \centering
    \begin{minipage}{0.48\textwidth}
        \centering
        \includegraphics[width=\textwidth]{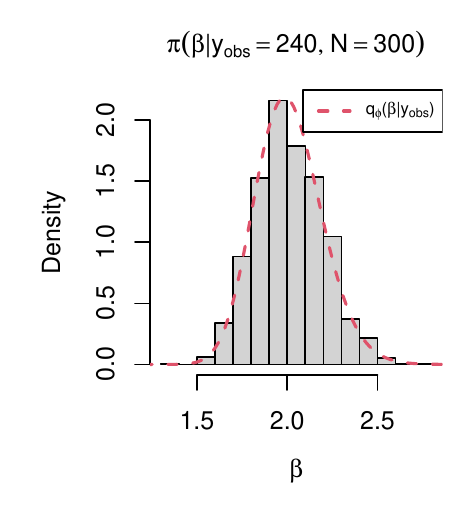}
    \end{minipage}
    \hfill
    \begin{minipage}{0.48\textwidth}
        \centering
        \includegraphics[width=\textwidth]{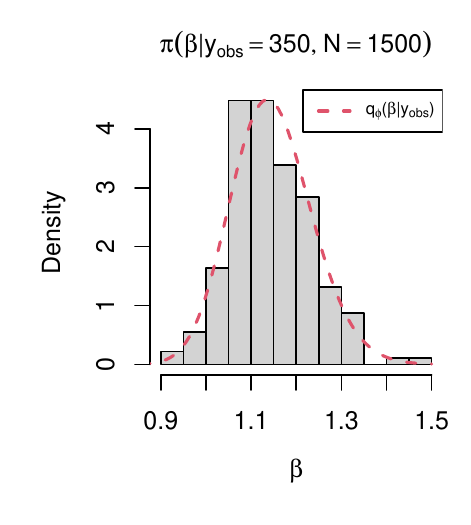}
    \end{minipage}
    \caption{Reference and NPE posterior distributions for the infection rate $\beta$ at two population sizes not included in the training data. Left: $y_{\mathrm{obs}} = 240$, $N = 300$, a major outbreak in a moderate-sized population. Right: $y_{\mathrm{obs}} = 350$, $N = 1500$, a minor outbreak in a large  population. In each case the histogram shows the reference posterior $\pi(\beta|y_{\mathrm{obs}}, N)$ obtained via exact-matching ABC, and the dashed curve shows the logNormal approximation $q_\phi(\beta|y_{\mathrm{obs}}, N)$ produced by the variable-$N$ network trained on populations of size $N \in \{100, 200, 500, 1000, 2000\}$.}
    \label{fig:varN_interpolation}
\end{figure}

Figure~\ref{fig:varN_interpolation} compares the NPE posterior with the reference posterior obtained via exact-matching ABC for each case. The agreement is close in both settings, indicating that the network has learned a sufficiently smooth function of $(y/N, N)$ to generalise reliably between training population sizes. The computational contrast between the two approaches is particularly stark for the larger dataset: obtaining the reference posterior for $N = 1500$ via exact-matching ABC required $10^7$ forward simulations, of which only 183 produced a final size of exactly $y_{\mathrm{obs}} = 350$, yielding an acceptance rate of under $0.002\%$. The NPE posterior, by contrast, was obtained from a single forward pass through the trained network with no additional simulation whatsoever.

\subsection{Abakaliki Smallpox Outbreak}\label{sec:abakaliki}
\begin{figure}[htpb]
    \centering
    \includegraphics[width=0.6\linewidth]{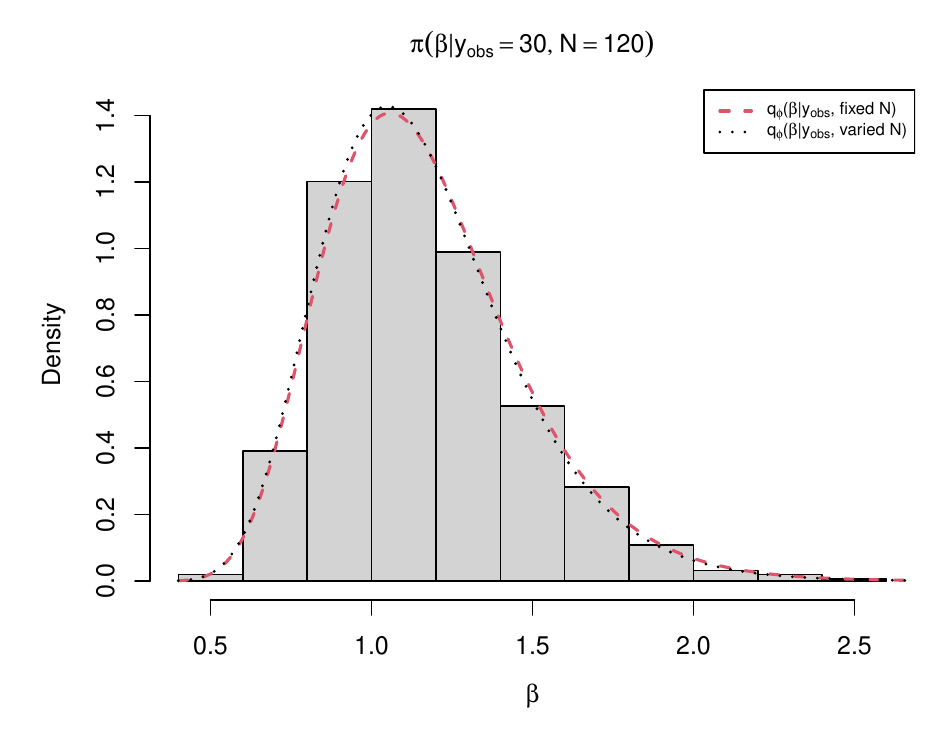}
    \caption{Posterior distributions for the infection rate $\beta$ for the Abakaliki smallpox data ($y_{\mathrm{obs}} = 30$, $N = 120$). The histogram shows the reference posterior $\pi(\beta|y_{\mathrm{obs}}, N)$ obtained via exact-matching ABC. The two curves show the logNormal NPE approximations: one from a network trained on $10^6$ simulations at the fixed population size $N = 120$ (fixed-$N$ NPE), and one from the network trained on the pooled dataset of $5 \times 10^6$ simulations across five population sizes ($N \in \{100, 200, 500, 1000, 2000\}$), with $N = 120$ supplied as an additional input at inference time (variable-$N$ NPE). } %The variable-$N$ network was not trained at $N = 120$; its approximation relies on interpolation between the two nearest training values, $N = 100$ and $N = 200$.}
    \label{fig:abakaliki-posteriors}
\end{figure}

We apply the NPE framework to the Abakaliki smallpox dataset \citep[see, e.g.,][]{ONeill_2009}, which records $y_{\mathrm{obs}} = 30$ cases in a closed community of $N = 120$ individuals and has served as a standard benchmark for Bayesian inference in epidemic models. We assign $\beta$ an $\mathrm{Exp}(1)$ prior and assume an $\mathrm{Exp}(1)$ infectious period, following the setup of Section \ref{sec:npe_homSIR_fixed_N}. We compare three posterior approximations: the true posterior obtained via exact-matching ABC; an NPE trained specifically at $N = 120$ using $10^6$ simulated parameter--data pairs (fixed-$N$ NPE); and the variable-$N$ NPE trained on the pooled dataset of $5 \times 10^6$ pairs across five population sizes, with $N = 120$ passed as an additional input at inference time. Crucially, $N = 120$ was not included among the training population sizes of the variable-$N$ network, so its posterior approximation relies entirely on interpolation between the two nearest training values, $N = 100$ and $N = 200$. Figure~\ref{fig:abakaliki-posteriors} shows the three posteriors. The fixed-$N$ NPE agrees closely with the reference posterior, consistent with the synthetic data results of Section~\ref{sec:npe_homSIR_fixed_N}. The variable-$N$ NPE is in similarly close agreement with both, demonstrating that the network generalises accurately to a population size outside its training set.

\newpage
\section{Household SIR model} \label{sec:household-SIR}
\subsection{Experimental setup}
We now turn to the more complex setting of the household SIR model introduced in Section~\ref{sec:two-level-def}. Unlike the homogeneously mixing model, which involves a single scalar parameter inferred from a one-dimensional
observation, the household model requires joint inference on two transmission rates from a multivariate count vector summarising the household outbreak size distribution. The inferential objective is to estimate the global and local transmission rates, $\lambda_G$ and $\lambda_L$. We assume that the infectious period is constant and of length one and assign independent Exponential prior distributions with rate 1 to both $\lambda_G$ and $\lambda_L$.

The observed data in this setting consist of the final-size distribution within each household size class: for households of size $k$, we record the vector 
$\mathbf{y}_k = (y_{k,0}, y_{k,1}, \ldots, y_{k,k})$, where $y_{k,j}$ 
denotes the number of households of size $k$, for $k=1, \ldots, K$, and in which exactly $j$ members  were infected over the course of the epidemic. The full observed summary statistic is the concatenation of these vectors across all household sizes present in the population. This summary is sufficient for inference under the independence approximation \citep{Addy_etal_1991} and has been used as the basis for likelihood-free inference in related work \citep{Neal_2012}.

Since the model involves two parameters, we approximate the joint posterior $\pi(\lambda_G, \lambda_L | \mathbf{y})$ by a bivariate logNormal distribution:
\begin{equation}
    q_\phi(\lambda_G, \lambda_L | \mathbf{y}) \equiv 
    \text{logNormal}_2\!\left(\boldsymbol{\mu}_\phi(\mathbf{y}),\, 
    \Sigma_\phi(\mathbf{y})\right),
\end{equation}
where $\boldsymbol{\mu}_\phi(\mathbf{y}) \in \mathbb{R}^2$ is the mean vector on the log scale and $\Sigma_\phi(\mathbf{y})$ is a $2 \times 2$ covariance matrix, both parameterised as functions of the data via the neural network weights $\phi$. The covariance matrix is parameterised in terms of two marginal standard deviations $s_1(\mathbf{y}), s_2(\mathbf{y}) > 0$ and a correlation coefficient $\rho(\mathbf{y}) \in (-1, 1)$, with positive definiteness guaranteed by construction via the Cholesky factorisation. The neural network therefore outputs five scalar quantities for each input $\mathbf{y}$: the two log-scale means, the two marginal standard deviations, and the correlation. This choice ensures support on the positive quadrant and provides a flexible yet parsimonious approximation to posterior distributions that are typically unimodal and moderately concentrated in this setting. For all household model experiments in this Section, the network uses three hidden layers of 128 units each with ReLU activations, trained with the Adam optimiser at a learning rate of $10^{-3}$, a batch size of 1024, and 200 epochs. 

To assess the accuracy of the NPE approximation, we compare against a reference posterior obtained via rejection ABC. When the population consists of households of small and equal size (e.g.\ $d = 2$), exact matching on the household outbreak size distribution is computationally feasible; since the data are discrete and representable as a multivariate count vector, this yields an unbiased approximation to the true posterior. For larger or more heterogeneous household structures, exact matching becomes impractical due to the low probability of exact agreement between simulated and observed count vectors. In these cases, we instead employ rejection ABC with an $\ell_1$ discrepancy, accepting parameter draws for which $\sum_{k,j} |y_{k,j}^{\text{sim}} - y_{k,j}^{\text{obs}}| \leq \varepsilon$, with the tolerance $\varepsilon$ chosen as small as practically feasible given available computational resources and reported alongside the corresponding results. For real outbreak datasets, we additionally compare our posteriors against estimates reported in the existing literature.

\subsection{Synthetic data experiment} \label{sec:households_equal_size}
We begin with a population of $H = 100$ households each of size $d = 2$, giving a total population of $N = 200$ individuals. In this setting, the observed data $\mathbf{y} = (y_{2,0}, y_{2,1}, y_{2,2})$ record the number
of households in which zero, one, or both individuals were infected, respectively, and satisfy $y_{2,0} + y_{2,1} + y_{2,2} = 100$. The data therefore lie on a two-dimensional simplex, and the complete set of admissible count vectors is finite and easily enumerated. This makes the setting particularly transparent: exact matching ABC is straightforward to implement and provides reliable reference posteriors. In principle the diagnostic approach of Section~\ref{sec:npe_homSIR_fixed_N} could be extended here, since every admissible observation can be enumerated; however, visualising five learned network outputs as functions of a two-dimensional input is considerably less tractable than the scalar case, and we instead assess accuracy through direct posterior comparisons for representative datasets.

\begin{figure}[htbp]
    \centering
     {\large \textbf{Major outbreak}}\\[4pt]
    \begin{minipage}{0.49\linewidth}
        \centering
        \includegraphics[width=\linewidth]{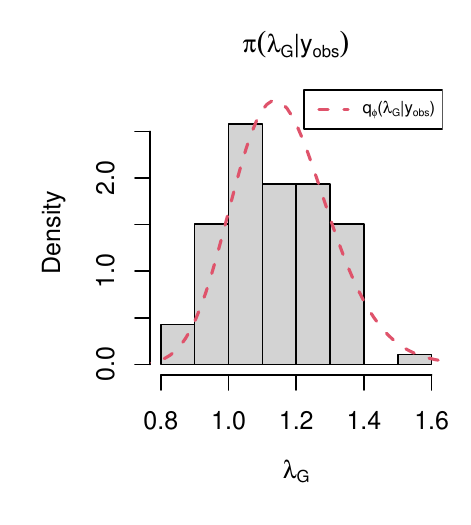}
    \end{minipage}
    \hfill
    \begin{minipage}{0.49\linewidth}
        \centering
        \includegraphics[width=\linewidth]{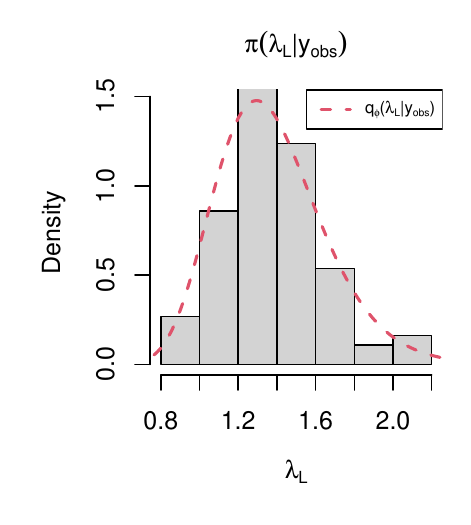}
    \end{minipage}
    \vspace{10pt}
     {\large \textbf{Minor outbreak}}\\[4pt]
    \begin{minipage}{0.49\linewidth}
        \centering
        \includegraphics[width=\linewidth]{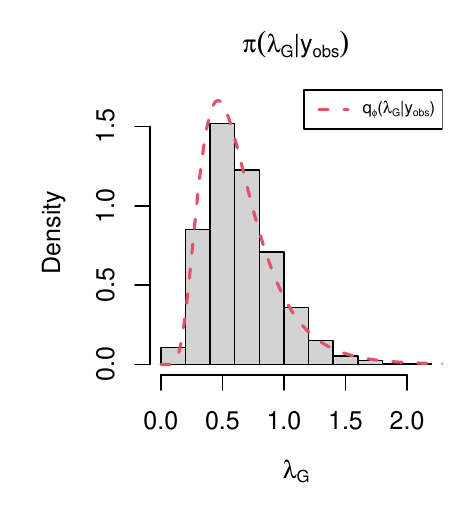}
    \end{minipage}
    \hfill
    \begin{minipage}{0.49\linewidth}
        \centering
        \includegraphics[width=\linewidth]{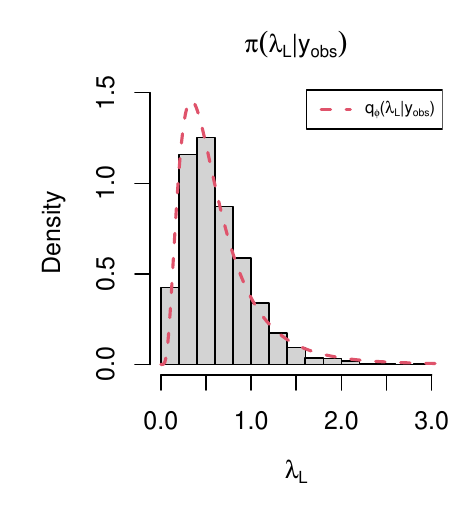}
    \end{minipage}
    \caption{Reference and NPE marginal posterior distributions of $\lambda_G$ and $\lambda_L$ for two observed datasets: $\mathbf{y} = (18, 12, 70)$ (top, major outbreak) and $\mathbf{y} = (95, 3, 2)$ (bottom, minor
    outbreak). The close agreement illustrates the accuracy of the bivariate logNormal neural posterior estimator.}
    \label{fig:household-SIR-H-200-same-size-2}
    
\end{figure}

We consider two synthetic datasets representing contrasting transmission regimes: a major outbreak, $\mathbf{y} = (18, 12, 70)$, in which the majority of households experienced full infection, and a minor outbreak, $\mathbf{y} = (95, 3, 2)$, in which most households escaped entirely. For each dataset, we compare the NPE marginal posterior approximations for $\lambda_G$ and $\lambda_L$ against reference posteriors obtained via exact matching ABC which yielded 93 and 2,585 accepted samples respectively from $10^6$ simulations. Figure~\ref{fig:household-SIR-H-200-same-size-2} shows that the NPE approximation closely matches the reference in both cases, demonstrating that the bivariate logNormal estimator accurately captures the conditional posterior structure across qualitatively different outbreak scenarios.

\subsection{Real outbreak data} \label{sec:household-real-data}
We apply NPE to two influenza datasets from household transmission studies in Seattle, Washington, available in \citet{longini1982} and previously analysed by \citet{Neal_2012} and \citet{ClancyOneill_2007}. The datasets record the distribution of household outbreak sizes for an influenza B outbreak (1975--1976) and an influenza A (H1N1) outbreak (1978--1979); see Table~\ref{tab:seattle-influenza}. The two outbreaks differ both in their household size structure---influenza A involves households of up to size 3, whereas influenza B includes households of up to size 5---and in their overall infection levels. In the influenza A outbreak, 97 out of 177 individuals were infected, indicating a substantially more infectious episode than influenza B, in which 56 out of 259 individuals were infected.

\begin{table}[htbp]
\centering
\caption{Observed distributions of household infection sizes during the influenza B (1975--1976) and influenza A (H1N1) (1978--1979) epidemics in Seattle, Washington. Entries give the number of households of each size with
the corresponding number of infected individuals. Data from \citet{longini1982}.}\label{tab:seattle-influenza}
\begin{tabular}{lrrrrr|rrrrr}
\toprule
 & \multicolumn{5}{c|}{Influenza B, 1975--1976} 
 & \multicolumn{5}{c}{Influenza A, 1978--1979} \\
No.\ infected & \multicolumn{5}{c|}{Household size} 
              & \multicolumn{5}{c}{Household size} \\
             & 1 & 2 & 3 & 4 & 5 & 1 & 2 & 3 & & \\
\midrule
0 & 9  & 12 & 18 & 9 & 4 & 15 & 12 & 4  \\
1 & 1  & 6  & 6  & 4 & 3 & 11 & 17 & 4  \\
2 &    & 2  & 3  & 4 & 0 &    & 21 & 4  \\
3 &    &    & 1  & 3 & 2 &    &    & 5  \\
4 &    &    &    & 0 & 0 &    &    &    \\
5 &    &    &    &   & 0 &    &    &    \\
\midrule
Total & 10 & 20 & 28 & 20 & 9 & 26 & 50 & 17 \\
\bottomrule
\end{tabular}
\end{table}

The two datasets are used to illustrate two distinct NPE deployment strategies, chosen to mirror the progression from fixed to variable population size in Section~\ref{sec:npe_homSIR_variable_N}. The influenza A outbreak is used to demonstrate the fixed-structure approach, in which the network is trained on a population matching the observed household composition exactly --- the most direct application of NPE. The influenza B outbreak then motivates the variable-structure approach: a fixed-structure network trained to match one specific population offers limited amortisation benefits when household compositions vary across settings, and we therefore train a single network across a range of plausible structures, without retraining for each new dataset encountered.

For the influenza A outbreak, the network is trained on simulated epidemics from a population whose with $(H_1, H_2, H_3) = (26, 50, 17)$ households of sizes 1, 2, and 3 respectively. The network input is the concatenated vector of {\em normalised} household final-size counts, 
$$ 
\left(
        \frac{y_{1,0}}{H_1}, \frac{y_{1,1}}{H_1},\quad
        \frac{y_{2,0}}{H_2}, \frac{y_{2,1}}{H_2}, \frac{y_{2,2}}{H_2},\quad
        \frac{y_{3,0}}{H_3}, \frac{y_{3,1}}{H_3}, \frac{y_{3,2}}{H_3}, \frac{y_{3,3}}{H_3}
\right)        
        \in [0,1]^{9},
$$
where each block is divided by the number of households of that size, giving a 9-dimensional input. This normalisation converts raw counts into empirical proportions within each size class, making the summary statistic comparable across simulations regardless of $H_k$, and ensuring that all inputs to the network lie in $[0,1]$. Since the population structure is fixed, $H_1, H_2, H_3$ carry no information and are not included as network inputs. The network therefore offers amortised inference for any outbreak outcome arising from populations with this specific household size distribution. 

For the influenza B outbreak, at each iteration of the training procedure, the household counts $H_1, \ldots, H_5$ are themselves drawn from a distribution over plausible population structures, so that the network sees a wide range of configurations during training. Specifically, at each training iteration the household composition $(H_1, \ldots, H_5)$ is drawn from a $\mathrm{Multinomial}\!\left(H,\, \hat{p}\right)$ distribution, where $H = 87$ is the total number of observed households and 
$\hat{p} = (H_1, \ldots, H_5)/H$ are the empirical proportions 
from the influenza B dataset. This keeps the total number of households 
fixed while randomising the composition across size classes, exposing the 
network to a range of population structures centred on the observed one. This requires extending the input representation to make the population structure explicit to the network. The observed household composition for the influenza B dataset is $(H_1, H_2, H_3, H_4, H_5) = (10, 20, 28, 20, 9)$  and the input vector is now
$$
    \boldsymbol{y} = 
    \left(
        \frac{y_{1,0}}{H_1}, \frac{y_{1,1}}{H_1},\;
        \ldots,\;
        \frac{y_{5,0}}{H_5}, \ldots, \frac{y_{5,5}}{H_5},\;
        \frac{H_1}{H}, \frac{H_2}{H}, \ldots, \frac{H_5}{H}
    \right) \in [0,1]^{25},
$$
where $H = \sum_{k=1}^{5} H_k$ is the total number of households.  The first 20 entries are the normalised final-size counts as for the influenza A data. The additional five entries encode the  population structure as \emph{proportions} of households of each size, rather than as raw counts.  This yields a single trained network capable of performing amortised inference across a family of household populations without retraining, at the cost of a harder learning problem. Results are reported in terms of the avoidance probabilities 
$q_G = \exp(-\lambda_G Q/N)$ and $q_L = \exp(-\lambda_L)$, following \citet{Neal_2012}. Here $Q$ denotes the final size of the epidemic, so $q_G$ is the probability that a susceptible individual avoids all global infectious contacts over the course of the epidemic, and $q_L$ is the probability of avoiding infection from a single local infective in their household, 
both under a constant infectious period of unit length.

\subsubsection*{Influenza A (H1N1), 1978--1979}

Table~\ref{tab:seattle-posteriors} reports posterior summary statistics for $(q_G, q_L)$ alongside the corresponding estimates from \citet{Neal_2012} and \citet{ClancyOneill_2007}. The NPE posterior means are 0.537 for $q_G$ and 0.701 for $q_L$, in close agreement with both references. The NPE credible intervals are marginally narrower than those of \citet{Neal_2012}, particularly for $q_L$. This is consistent with a known limitation of ABC-based methods when applied to multivariate count data: when exact matching ($\varepsilon_1 = \varepsilon_2 = 0$) is used, the coupled ABC algorithm of \citet{Neal_2012} yields only 32 accepted simulations for the influenza A data, which is insufficient for reliable posterior inference. Increasing the tolerance to $(\varepsilon_1, \varepsilon_2) = (12, 2)$ raises the number of accepted simulations to over 22,000, but at the cost of introducing additional posterior uncertainty: the accepted simulations are no longer draws from the exact posterior but from a smoothed approximation $\pi(\theta|\mathbf{y}^{\text{obs}}, \varepsilon)$, which inflates posterior variance relative to the true posterior. NPE sidesteps this trade-off entirely: the trained network produces a smooth approximation to the posterior without discarding any simulations or relaxing a tolerance criterion, and its posterior summaries are consequently not subject to this source of variance inflation. 

The somewhat tighter credible intervals obtained by NPE are consistent with the expected effect of tolerance inflation in ABC, which smooths the posterior and inflates variance relative to the true posterior. However, we note that the logNormal family itself introduces approximation error, and the two effects cannot be cleanly separated at this sample size. The close agreement in posterior means across all three methods, and the consistency of the NPE intervals with those of \citet{ClancyOneill_2007}, suggest that neither source of error is substantial here, but we caution against over-interpreting small differences in posterior spread between methods. Figure~\ref{fig:seattle-joint-posteriors} (left) shows the joint NPE posterior of $(q_G, q_L)$, which reveals a moderate positive correlation between the two avoidance probabilities.

\subsubsection*{Influenza B, 1975--1976}

\begin{figure}[htbp]
    \centering
    \begin{minipage}{0.49\linewidth}
        \centering
        \includegraphics[width=\linewidth]{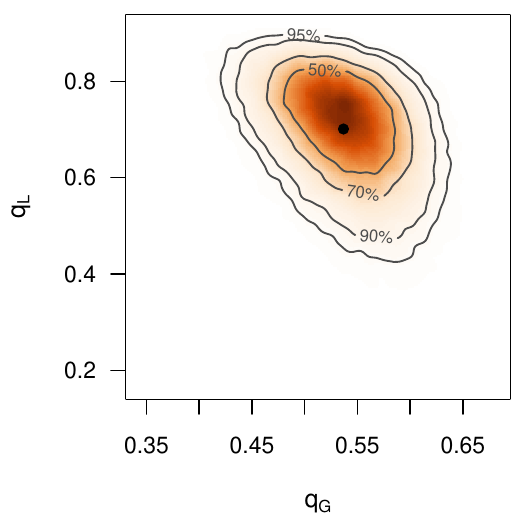}
    \end{minipage}
    \hfill
    \begin{minipage}{0.49\linewidth}
        \centering
        \includegraphics[width=\linewidth]{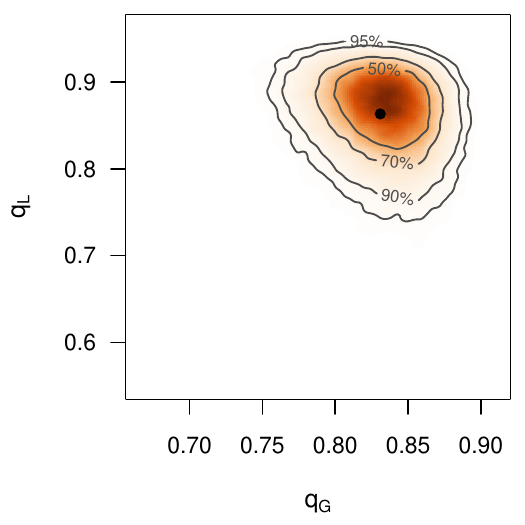}
    \end{minipage}
    \caption{Joint NPE posterior distributions of the avoidance probabilities $(q_G, q_L)$ for the influenza A (H1N1) 1978--1979 outbreak (left) and the influenza B 1975--1976 outbreak (right), both from Seattle, Washington. Filled surfaces show kernel density estimates of the joint posterior obtained from 100,000 NPE samples, with darker shading indicating higher posterior density. Contour lines correspond to the 50\%, 70\%, 90\%, and 95\% highest density regions. Filled circles mark the posterior means. The two posteriors reflect the markedly different transmission intensities of the two outbreaks, with the influenza A posterior concentrated at
    substantially lower avoidance probabilities and exhibiting greater posterior uncertainty.}
    \label{fig:seattle-joint-posteriors}
\end{figure}

Table~\ref{tab:seattle-posteriors} reports posterior summary statistics for $(q_G, q_L)$ alongside the corresponding estimates from \citet{Neal_2012} and \citet{ClancyOneill_2007}. The NPE posterior means are 0.831 for $q_G$ and 0.863 for $q_L$, both in close agreement with the published estimates. These values reflect the considerably milder nature of this outbreak relative to influenza A: a susceptible individual has approximately an 83\% chance of avoiding a global infectious contact and an 86\% chance of avoiding infection from a single local infective. The NPE credible intervals are again marginally narrower than those of 
\citet{Neal_2012}, consistent with the discussion of tolerance inflation 
and approximation error above. Notably, the posterior standard deviations for both parameters are substantially smaller than those obtained for influenza A, reflecting the greater informativeness of the larger influenza B dataset and the more concentrated posterior arising from the higher avoidance probabilities.
Figure~\ref{fig:seattle-joint-posteriors} (right) shows the joint NPE posterior of $(q_G, q_L)$.

\begin{table}[htbp]
\centering
\caption{Posterior summary statistics for the avoidance probabilities $(q_G, q_L)$ for the influenza A (H1N1) 1978--1979 and influenza B 1975--1976 Seattle outbreaks. Equal-tailed 95\% credible intervals are reported. For \citet{ClancyOneill_2007}, only posterior means are available from the published analysis.}
\label{tab:seattle-posteriors}
\begin{tabular}{lcccc}
\toprule
 & \multicolumn{2}{c}{$q_G$} 
 & \multicolumn{2}{c}{$q_L$} \\
\cmidrule(lr){2-3}\cmidrule(lr){4-5}
Method & Mean (sd) & 95\% CI & Mean (sd) & 95\% CI \\
\midrule
\multicolumn{5}{l}{\textit{Influenza A, 1978--1979}} \\
\midrule
NPE 
    & 0.537 (0.043) & (0.448, 0.618) 
    & 0.701 (0.099) & (0.475, 0.858) \\
\citet{Neal_2012} 
    & 0.532 (0.048) & (0.437, 0.623) 
    & 0.686 (0.106) & (0.486, 0.898) \\
\citet{ClancyOneill_2007} 
    & 0.54 \phantom{(0.000)} & --- 
    & 0.69 \phantom{(0.000)} & --- \\
\midrule
\multicolumn{5}{l}{\textit{Influenza B, 1975--1976}} \\
\midrule
NPE 
    & 0.831 (0.028) & (0.771, 0.880) 
    & 0.863 (0.044) & (0.761, 0.931) \\
\citet{Neal_2012} 
    & 0.833 (0.031) & (0.767, 0.888) 
    & 0.854 (0.062) & (0.725, 0.969) \\
\citet{ClancyOneill_2007} 
    & 0.83 \phantom{(0.000)} & --- 
    & 0.87 \phantom{(0.000)} & --- \\
\bottomrule
\end{tabular}
\end{table}

\section{Discussion}\label{sec:discussion}

We have presented a neural posterior estimation framework for Bayesian inference in stochastic SIR epidemic models observed through final outcome data, and demonstrated its application to two model classes of increasing complexity: a homogeneously mixing SIR model with a scalar transmission parameter, and a two-level household SIR model requiring joint inference on global and local transmission rates. In both settings, the NPE approximation closely matched reference posteriors obtained via rejection sampling (through exact-matching ABC) and, on real outbreak datasets, produced results in close agreement with those reported in the existing literature. To the best of our knowledge, this is the first application of NPE to stochastic epidemic models in the final outcome observation setting. The results are directly relevant to the retrospective outbreak investigations and household transmission studies described in the Introduction (see Section \ref{sec:intro}).

The methodology has several practically appealing features. \citet{DemirisONeill_2005} note that, while their data-augmentation MCMC methods work well in practice, datasets involving very large numbers of cases --- thousands as opposed to hundreds --- can require days rather than hours to analyse, though they add that such large outbreaks are uncommon. The NPE framework does not share this limitation. Training is carried out offline using simulated data, and once the network has been trained, posterior 
approximations are obtained via a single forward pass through the network, requiring a fraction of a second regardless of population size. The variable-$N$ experiments in Section~\ref{sec:npe_homSIR_variable_N} demonstrate accurate posterior inference at population sizes up to $N = 2000$, with reliable interpolation to unseen population sizes, at negligible additional computational cost. This scalability is a direct consequence of the simulation-based nature of the approach: inferential complexity does not grow with population size, only the cost of forward simulation does, and for SIR-type models forward simulation remains cheap even at large $N$.

Using the Sellke construction, generating $10^6$ training pairs requires seconds for populations of size $N = 100$--$200$ and a few minutes at $N = 2000$. Network training, implemented in PyTorch, ranged from a few minutes for fixed-$N$ networks to approximately two hours for the variable-$N$ and household models, running on a standard desktop machine with no specialist hardware. This training cost is a one-time overhead: once the network is trained, it can be reused for any number of future datasets at negligible additional cost, so the investment pays off rapidly whenever repeated inference is anticipated.

The presented framework is also flexible and relatively straightforward to extend: as demonstrated in the variable-$N$ and variable-structure experiments, additional contextual information such as population size or household composition can be incorporated by simply augmenting the network input, without redesigning the underlying algorithm or deriving new computational procedures. Implementation requires only the ability to simulate from the model and access to standard neural network training tools, both of which are readily available in modern statistical computing environments. The framework also extends naturally to settings of partial observation, where final outcome data are available for only a fraction of the population, as arises for example in seroprevalence surveys. Within the NPE training procedure, this requires only an additional simulation step in which the observed data are obtained by subsampling from the simulated epidemic outcome; the sampling fraction can then be passed as a further network input, allowing a single trained network to handle varying levels of observation without any modification to the underlying algorithm.

As with all simulation-based inference methods, NPE requires that training simulations be generated from a prior distribution, and the choice of prior therefore influences not only the posterior but also the efficiency of the training procedure. For epidemic models this requires some care. Epidemic processes exhibit threshold behaviour: when the reproduction number $R_0$ is close to or below 1, most simulated epidemics will die out after infecting very few individuals, while a highly diffuse prior that places substantial mass well above the threshold will generate training data dominated by near-total infection. In either case, relatively few simulations fall in the epidemiologically informative regime (i.e. outbreaks of intermediate size that carry most of the information about transmission parameters) and the network may learn poorly in precisely the region that matters for inference. The Exponential priors with rate 1 used throughout this paper were chosen to concentrate training simulations in this epidemiologically informative range, i.e. outbreaks neither too small to be uninformative nor so large as to reflect unrealistically high transmission, while remaining consistent with standard modelling practice. In settings where a less informative prior is required, Sequential NPE \citep[e.g.][]{Cranmer_etal_2020} can be employed which iteratively refines the proposal distribution for training simulations, progressively focusing computation on the posterior-relevant region without committing to an informative prior from the outset. %Alternatively, a small number of pilot simulations under a diffuse prior can be used to identify the region of parameter space that generates epidemiologically plausible outcomes, and a more targeted training proposal constructed accordingly. Either approach allows the practitioner to retain broad prior uncertainty at the inference stage while ensuring that training simulations are concentrated where the likelihood is informative.

The use of logNormal families as posterior approximations proved effective throughout: the univariate and bivariate logNormal estimators accurately captured the conditional posterior structure across all models and datasets considered, reflecting the fact that posteriors for these models are typically unimodal and moderately concentrated over a positive parameter space. For more complex epidemic models, such as multitype or spatially structured models with higher-dimensional parameter spaces, the posterior may exhibit multimodality, stronger nonlinear dependencies, or heavier tails, and more flexible density estimators would be the natural choice. The NPE framework accommodates this directly: normalising flows \citep[e.g.][]{lueckmann2021benchmarking}, neural spline flows \citep{durkan2019neural}, and Gaussian mixture networks \citep{bishop1994mixture} can all serve as drop-in replacements for the logNormal family, retaining the same simulation-based training procedure while placing fewer restrictions on the shape of the approximating distribution. Extending the present methodology in this direction, with application to more complex structured population models, is the subject of current work.

Finally, the present paper has focused on final outcome data. An important and natural extension is to the setting where individual-level temporal data are available, such as removal times or sequences of symptom onsets. Such data are substantially more informative about model parameters but introduce additional inferential challenges: the observed data are higher-dimensional, 
requiring either careful construction of informative summary statistics or architectures capable of handling variable-length sequences directly; and posterior distributions may exhibit more complex geometry, including stronger parameter dependencies than those encountered in the final outcome data setting. Preliminary investigations suggest that the NPE framework extends naturally to this setting, and a full treatment of temporal epidemic data using NPE is the subject of ongoing work. %Code for training all NPE models presented in this paper is publicly available at \url{https://github.com/kypraios/[repository]}.

\section*{Declaration of generative AI and AI-assisted technologies in the manuscript preparation process}
During the preparation of this work the author used {\tt Claude.ai} to assist with editorial review, including improving clarity and structure of the text. After using this tool/service, the author reviewed and edited the content as needed and takes full responsibility for the content of the published article.

\section*{Acknowledgments}
The author is grateful to the Leverhulme Trust for support through a Research Fellowship [RF-2025-377] and to the National Institute for Health and Care Research (NIHR) through the Better Methods Better Research programme [NIHR174268]. The author would like to thank Professors Frank Ball and Peter Neal for helpful discussions on efficient simulation of final outcome data from household SIR models. The views expressed are those of the author and not necessarily those of the NIHR or the Department of Health and Social Care.

\bibliographystyle{apalike}
\bibliography{refs}

\newpage
\section*{Appendix}
\appendix
\renewcommand{\thefigure}{S\arabic{figure}}
\setcounter{figure}{0}

% ============================================================
% Figure: learned vs empirical log-mean mu(y, N)
% ============================================================
\begin{figure}[ht]
    \centering

    % --- Row 1 ---
    \begin{minipage}[t]{0.45\textwidth}
        \centering
        \includegraphics[width=\textwidth]{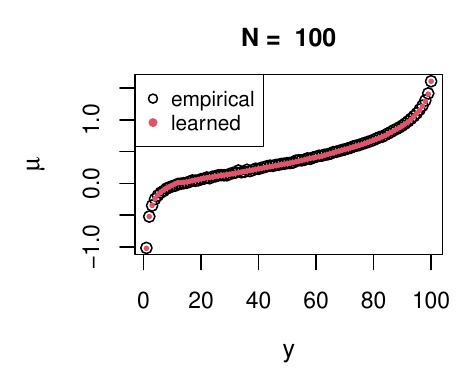}
    \end{minipage}
    \hfill
    \begin{minipage}[t]{0.45\textwidth}
        \centering
        \includegraphics[width=\textwidth]{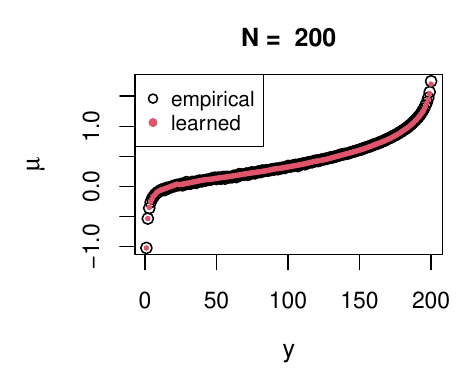}
    \end{minipage}

    \vspace{-2em}

    % --- Row 2 ---
    \begin{minipage}[t]{0.45\textwidth}
        \centering
        \includegraphics[width=\textwidth]{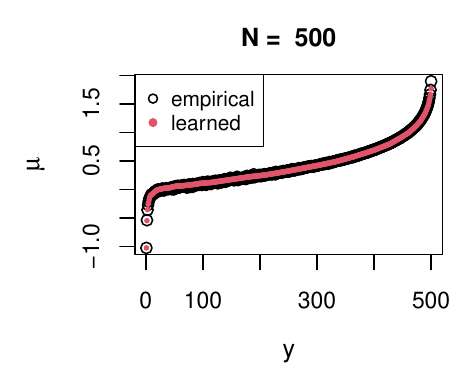}
    \end{minipage}
    \hfill
    \begin{minipage}[t]{0.45\textwidth}
        \centering
        \includegraphics[width=\textwidth]{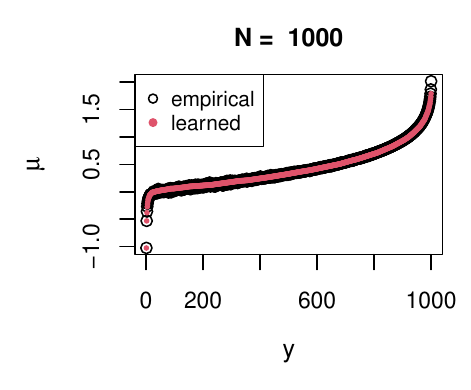}
    \end{minipage}
    
    \vspace{-2em}

    % --- Row 3 ---
    \begin{minipage}[t]{0.45\textwidth}
        \includegraphics[width=\textwidth]{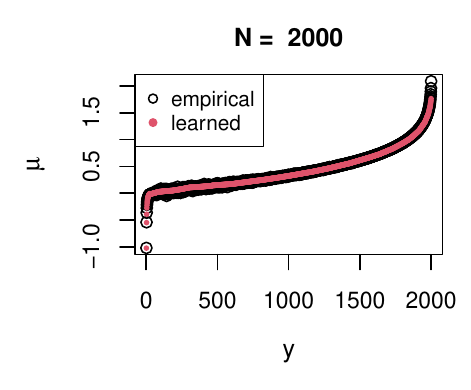}
    \end{minipage}

    \caption{Learned parameter $\mu_\phi(y, N)$ of the logNormal posterior approximation as a function of the final size $y$, shown separately for each of the five population sizes (100, 200, 500, 1000, 2000) used in training. Red solid circles show the outputs of the trained neural network; white circles correspond to empirical estimates obtained by fitting a logNormal distribution to simulated $\beta$ values by exact matching on $y$. The close agreement across all population sizes indicates that the single network has learned the conditional posterior structure for each $N$ simultaneously.}
    \label{fig:varN_mu}
\end{figure}

% ============================================================
% Figure: learned vs empirical log-mean mu(y, N)
% ============================================================
\begin{figure}[h]
    \centering

    % --- Row 1 ---
    \begin{minipage}[t]{0.48\textwidth}
        \centering
        \includegraphics[width=\textwidth]{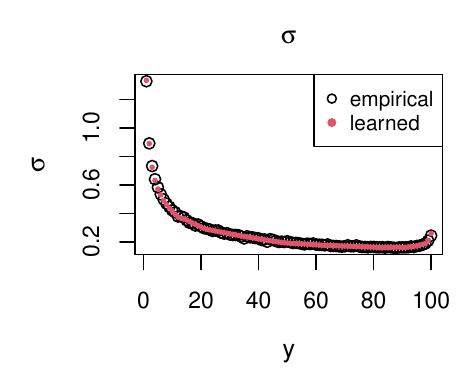}
    \end{minipage}
    \hfill
    \begin{minipage}[t]{0.48\textwidth}
        \centering
        \includegraphics[width=\textwidth]{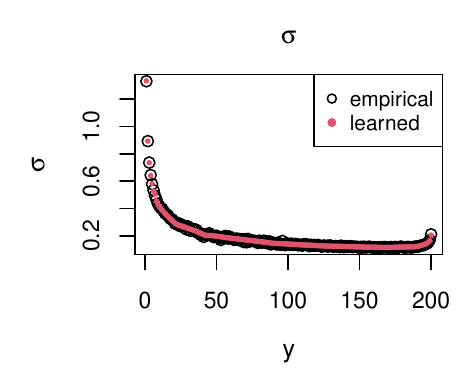}
    \end{minipage}

    \vspace{-2em}

    % --- Row 2 ---
    \begin{minipage}[t]{0.48\textwidth}
        \centering
        \includegraphics[width=\textwidth]{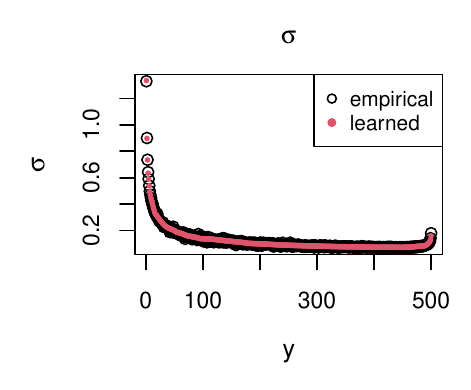}
    \end{minipage}
    \hfill
    \begin{minipage}[t]{0.48\textwidth}
        \centering
        \includegraphics[width=\textwidth]{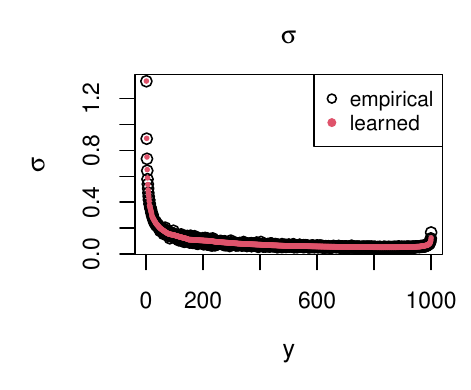}
    \end{minipage}
    
    \vspace{-2em}

    % --- Row 3 ---
    \begin{minipage}[t]{0.48\textwidth}
        \includegraphics[width=\textwidth]{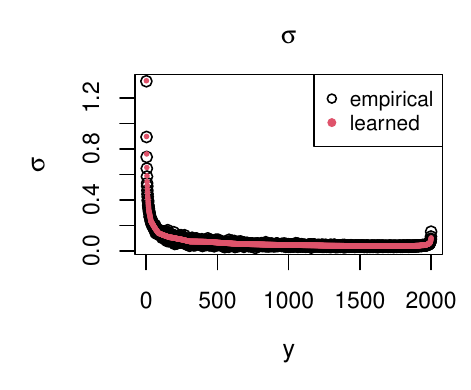}
    \end{minipage}

    \caption{Learned parameter $\sigma_\phi(y, N)$ of the logNormal posterior approximation as a function of the final size $y$, shown separately for each of the five population sizes (100, 200, 500, 1000, 2000) used in training. Red solid circles show the outputs of the trained neural network; white circles correspond to empirical estimates obtained by exact matching on $y$. The decrease in $\sigma_\phi(y, N)$ with increasing $N$, for a given proportion infected $y/N$, reflects the greater information content of larger outbreaks.}
    \label{fig:varN_sigma}
\end{figure}

\end{document}